\documentclass{PoS}
\usepackage{color,caption}
\usepackage[all]{xy}
\usepackage{relsize}
\usepackage{picture}
\usepackage{pifont,picture,multicol,amsfonts}
\usepackage{psfrag}
\usepackage{palatino}
\usepackage{amsfonts,bm,amssymb,euscript,array,cite}

\usepackage{tikz}
\newcommand*\circled[1]{\tikz[baseline=(char.base)]{\node[shape=circle,draw,minimum width=1cm,inner sep=2pt] (char) {$#1$};}}
\usepackage{amsmath,amsthm}
\usepackage[]{graphicx}
\usepackage[T1]{fontenc}
\usepackage{framed}

\title{Gauging as constraining: the universal \\ generalised
geometry action in two dimensions\thanks{Dedicated to the memory of Ioannis Bakas.}}
\ShortTitle{Universal gauge theory in two dimensions}

\author{\speaker{Athanasios Chatzistavrakidis}%
      \\
       Van Swinderen Institute for Particle Physics and 
Gravity, University of Groningen, \\ 
Nijenborgh 4, 9747 AG Groningen, The Netherlands\\
       E-mail: \email{a.chatzistavrakidis@rug.nl}}

\author{Andreas Deser\\
  Istituto Nationale di Fisica Nucleare - Sezione di Torino\\
  Via Pietro Giuria 1, 10125 Torino, Italy\\
        E-mail: \email{deser@to.infn.it}}

        \author{Larisa Jonke\\
        Division of Theoretical Physics, 
 Rudjer Bo$\check s$kovi\'c Institute, \\
 Bijeni$\check c$ka 54, 10000  Zagreb, Croatia\\
        E-mail: \email{larisa@irb.hr}}
        \author{Thomas Strobl\\ UMI CNRS2924 Instituto de Matem\'atica Pura e Aplicada (IMPA), \\ Estrada Dona Castorina 110, Rio de Janeiro, 22460-320, Brasil  \\ \emph{and} \\
        Institut Camille Jordan, Universit\'e Claude Bernard Lyon 1, \\
43 boulevard du 11 novembre 1918, 69622 Villeurbanne cedex, France\\
        E-mail: \email{thomasATimpa.br and stroblATmath.univ-lyon1.fr}}

  \abstract{One of the central concepts in modern theoretical physics, gauge symmetry, is typically realised by lifting a finite-dimensional global symmetry group of a given functional to an infinite-dimensional local one by extending the functional to include gauge fields. In this contribution we review the construction of gauged actions for two-dimensional sigma models, considering a more general notion to be gauged, namely that of a (possibly singular) foliation. In particular, the original action does not need to have any global symmetry for this purpose. Moreover, reformulating the ungauged theory by means of auxiliary 1-form fields taking values in the generalised tangent bundle over the target, all possible such gauge theories result from restriction of these fields to take values in (possibly small) Dirac structures. This turns all the remaining 1-form fields into gauge fields and leads to the presence of a local symmetry.  We recall all needed mathematical notions, those of (higher) Lie algebroids,  Courant algebroids, and Dirac structures.}

\FullConference{Corfu Summer Institute 2016 "School and Workshops on Elementary Particle Physics and Gravity"\\
                 31 August - 23 September, 2016\\
                 Corfu, Greece}

\begin{document}

\newcommand{\sfrac}[2]{{\textstyle\frac{#1}{#2}}}
\newcommand{\co}{{\cal O}}
\newcommand{\cross}{\ding{55}}

\newcommand{\be}{\begin{equation}} \newcommand{\ee}{\end{equation}}
\newcommand{\beq}{\begin{equation}} \newcommand{\eeq}{\end{equation}}
\newcommand{\beqa}{\begin{eqnarray}}
\newcommand{\eeqa}{\end{eqnarray}} \newcommand{\eq}[1]{(\ref{#1})}
 \def\bea{\begin{eqnarray}} \def\eea{\end{eqnarray}}
\def\obar{\overline}
\def\ss{\scriptsize}
\newcommand{\barr}{\begin{array}}
\newcommand{\earr}{\end{array}}
\newcommand{\ct}[1]{{\ss{\ki{#1}}}}
\newcommand{\vt}{\vspace{3pt}}

\newtheorem{theorem}{Theorem}
%

\def\a{\alpha} \def\da{{\dot\alpha}} \def\b{\beta}
\def\db{{\dot\beta}} \def\g{\gamma} \def\G{\Gamma}
\def\cdt{\dot\gamma} \def\d{\delta} \def\D{\Delta}
\def\ddt{\dot\delta} \def\e{\epsilon} \def\vare{\varepsilon}
\def\f{\phi} \def\F{\Phi} \def\vvf{\f} \def\h{\eta} \def\k{\kappa}
\def\l{\lambda} \def\L{\Lambda} \def\la{\lambda} \def\m{\mu}
\def\n{\nu} \def\o{\omega} \def\p{\pi} \def\P{\Pi} \def\r{\rho}
\def\s{\sigma} \def\S{\Sigma} \def\t{\tau} \def\th{\theta}
\def\Th{\Theta} \def\vth{\vartheta} \def\X{\Xeta} \def\z{\zeta}

\def\u{\upsilon}


\def\cA{{\cal A}} \def\cB{{\cal B}} \def\cC{{\cal C}} \def\cD{{\cal
D}} \def\cE{{\cal E}} \def\cF{{\cal F}} \def\cG{{\cal G}}
\def\cH{{\cal H}} \def\cI{{\cal I}} \def\cJ{{\cal J}} \def\cK{{\cal
K}} \def\cL{{\cal L}} \def\cM{{\cal M}} \def\cN{{\cal N}}
\def\cO{{\cal O}} \def\cP{{\cal P}} \def\cQ{{\cal Q}} \def\cR{{\cal
R}} \def\cS{{\cal S}} \def\cT{{\cal T}} \def\cU{{\cal U}}
\def\cV{{\cal V}} \def\cW{{\cal W}} \def\cX{{\cal X}} \def\cY{{\cal
Y}} \def\cZ{{\cal Z}}


\def\R{{\mathbb R}} \def\C{{\mathbb C}} \def\N{{\mathbb N}}
\def\Z{{\mathbb Z}} \def\one{\mbox{1 \kern-.59em {\rm l}}}

\def\mr{\mathfrak{r}} \def\mt{\mathfrak{t}}
\def\mg{\mathfrak{g}}
\def\mmu{\mathfrak{u}}
\def\msu{\mathfrak{su}}


\def\bit{\begin{itemize}} \def\eit{\end{itemize}} \def\Tr{\mbox{Tr}}

\def\({\left(} \def\){\right)} \def\tens{\otimes} \def\Pint{\int
\!\!\!\!\!\!P} \def\rep{representation } \def\reps{representations }
\def\und{\underline}

\def\im{\imath}

\def\tvp{\varphi}

\newcommand{\nn}{\nonumber}
\newcommand{\miso}{\frac{1}{2}}
\def\beq{\begin{equation}}
\def\eeq{\end{equation}}
\def\bea{\begin{eqnarray}}
\def\eea{\end{eqnarray}}

\newcommand{\fet}{\frac{1}{3}}
\newcommand{\fdt}{\frac{2}{3}}
\newcommand{\ftt}{\frac{4}{3}}
\def\w{\wedge}
\def\olra{\overleftrightarrow}
\def\la{\textcolor[rgb]{0.00,0.50,1.00}}
\def\ra{\textcolor[rgb]{0.50,0.25,0.25}}
\def\ki{\textcolor[rgb]{0.00,0.50,0.00}}
\def\bi{\begin{itemize}}
\def\ei{\end{itemize}}
\def\tc{\textcolor}
\def\sq{\rightsquigarrow}
\newcommand{\T}{\text{T}}
\newcommand{\M}{\text{M}}
\newcommand{\dd}{\mathrm{d}}
\newcommand{\mf}{\mathfrak}

\def\tb{\mc T\text{M}}
\def\mx{\mathfrak{X}}

\newcommand{\nitem}{\item[{\scriptsize\color{DarkBlue}\ding{68}}]}
\newcommand{\sitem}{\item[{\scriptsize\color{DarkBlue}\ding{95}}]}
\newcommand{\titem}{\item[{\scriptsize\color{DarkBlue}\ding{51}}]}
\newcommand{\ttitem}{\item[{\scriptsize\color{DarkBlue}\ding{52}}]}
\newcommand{\ssitem}{\item[{\scriptsize\color{DarkBlue}\ding{96}}]}
\newcommand{\nnitem}{\item[{\scriptsize\color{DarkBlue}\ding{104}}]}

\section{Introduction and Motivations}

Symmetry principles and gauge field theories are a cornerstone of modern theoretical physics. 
The central concept of \emph{gauging}, the construction of theories which have invariant Lagrangians under some local symmetry, 
is typically associated to Lie groups. When a Lie group acts on a manifold $M$, it 
generates a partition of $M$ into leaves forming a foliation ${\cal F}$. These leaves are the orbits of the group. 
Depending on the type of the group action, for example whether it is the adjoint of the group on itself or a free action on some other manifold, the resulting foliation may be  regular (all leaves have the same dimension---like in the case of a free action) or singular (the dimension of the leaves varies, for instance there is a fixed point---like the identity element w.r.t.~the adjoint action).  
However, the notion of a foliation is much more general and does not rely on a group structure. 
Thus it is natural to wonder whether one can construct 
gauge field theories without having at hand a group action, working directly at the level of foliations. 
Such a task was undertaken in our recent work \cite{Chatzistavrakidis:2016jci,Chatzistavrakidis:2016jfz}, building on \cite{withoutsymmetry, Kotov:2004wz,Kotov:2014dqa}. In this contribution 
we review and clarify the main ideas and some of the results found there, namely on the gauging of a singular foliation
in the case of 2D bosonic $\sigma$-models. 

Apart from the main motivation of exploring  local symmetries of field theories in full generality 
and contructing new types of gauge theories, 
there is a host of additional motivations for this work, mainly in the context of string theory. 
The study of the possible target spaces where closed strings can consistently propagate is a perpetually interesting theme. 
Wess-Zumino-Witten (WZW) models have a prominent position in this respect, being consistent string theories with target spaces being 
group manifolds $G$ \cite{Witten:1983ar}. Moreover, there exists a consistent gauging of the adjoint action of any subgroup 
$K\subset G$, as shown in \cite{Gawedzki:1988hq,Gawedzki:1988nj}, leading to gauged $\sigma$-models. But, in contrast to what one might think naively, namely that the theory is one with effectively a string propagating in a target manifold $G/ \mathrm{Ad}(K)$, this turns out to not be the case, there is an additional freezing of propagation in this target due to the presence of the gauge fields. In this context it is thus natural to ask the following questions: \emph{Can we describe string propagation in quotient spaces beyond the realm of 
group actions?}  And:  \emph{Can we do so with keeping the propagation space of the string modulo the gauge symmetries to coincide with the leaf space, i.e.~the gauge theory to not produce extra unnatural freezing such as in the $G/K$ WZW theory?} We answer these questions all essentially in the affirmative: Under fairly general conditions on the background data, in particular much more general ones than those of invariance conditions, gauging of a large class of  singular foliations is possible. Moreover, freezing can be avoided in general, although possibly not on a global level. We will review the answer to the first question in the present contribution, while for the second point we refer to  \cite{Chatzistavrakidis:2016jfz}; in that paper we showed in particular that for the $G/G$ model with $G=SU(2)$ such a gauging, namely one without an additional freezing, is in fact obstructed on all of $G$---while it can be realised on all of $G$ except for an arbitrarily small region. 

A further stringy motivation originates in non-Abelian T-duality, or more generally in Poisson-Lie (PL) T-duality \cite{Klimcik:1995ux}. 
One question then is how one can generalize the particular Drinfeld double structure and establish a generalization of PL T-duality 
beyond groups. An accompanying question refers to the relation to generalized geometry and its characteristic Courant algebroid 
structure, which is anyway a natural generalization of the Drinfeld double for Lie bialgebroids. As we will explain in the ensuing, such structures 
emerge naturally from the gauging procedure we will follow. 
Finally, a related motivation comes from the study of  the so-called ``non-geometric'' string backgrounds, which constitute novel types 
of geometries where closed strings propagate and exhibit attractive features for flux compactifications of string theory, such as 
improved stability properties for moduli fields. 
Worldsheet/worldvolume perspectives on such backgrounds were given in Refs.~\cite{Hull:2004in,Halmagyi:2008dr,Mylonas:2012pg,Chatzistavrakidis:2015vka}.

In this contribution we 
 focus on 2D bosonic $\s$-models at leading order in the string slope parameter, thus ignoring dilaton contributions. 
One of our main goals is to determine the conditions for existence of \emph{some} gauge extension of such models. 
 In other words, we intend to couple gauge fields $A$ to the theory, valued in \emph{some} ``gauge'' bundle  
and for appropriate gauge transformations, determine the right-hand side of the Lie derivatives 
${{\cal L}_{\rho}g=\dots}\quad \mbox{and} \quad {{\cal L}_{\rho}B=\dots}\quad \mbox{or}
\quad {{\cal L}_{\rho}H=\dots}$
such that the theory is gauge invariant.
Before delving into more details, let us offer a prospectus of our main results
\bi 
\item The right-hand side of the invariance conditions need not be zero
and the gauging is controlled by two (curved, in general) connections $\nabla^{\pm}$.
\item There exists a \emph{universal gauge theory} with target $TM\oplus T^{\ast}M$.
\item In  certain ``maximal'' cases it generalises $G/G$ WZW and twisted or WZ Poisson $\sigma$-models, and $\nabla^{\pm}$ are then determined 
 in closed form.
\ei 
There are, however, also two original contributions in this article: One is a new example of gaugings without initial symmetry, extending the previously found ones 
with lower-dimensional target of \cite{Chatzistavrakidis:2015lga,Chatzistavrakidis:2016vsy} to an interesting new class of examples. The other one is a shift of perspective on 
the  gauging procedure of foliations, them coming from groups or not: Instead of considering the gauging procedure as an extension process of a given functional, we may
regard an equivalent, enlarged, universal one from which all the gaugings result upon constraining. This is interestingly close in spirit to the Hamiltonian framework, where a gauged theory 
results from an ungauged one by adding constraints on the phase space in a consistent way. Consistent usually means of the first class, and for two-dimensional theories this 
was seen to be intimately related to (small) Dirac structures in \cite{AlekseevStrobl}. The universal action point of view provides an interesting analog of this within the 
Lagrangian framework, and for the two-dimensional theories discussed in this paper, it is again the (small) Dirac structures which govern possible gaugings.

In order to explain how these results are obtained, it is useful to present and elucidate some basic mathematical notions (section 2). In Section 3 we discuss the gauging procedure, starting with groups and ending with foliations. In Section 4 we present the form of the universal action functional that controls all possible gaugings. We explain how the gauging results from constraining and relate the findings of the preceding section to generalised geometry. Sections 5 and 6 contain examples, in the form of topological and non-topological ones, respectively.

\newpage
\section{Going beyond Lie algebras and standard geometry}

We first recall that a Lie algebra $\mf g$ is a vector space together with an antisymmetric bilinear product denoted by a bracket $[\cdot,\cdot]$
such that the Jacobi identity
\begin{equation}
[a,[b,c]] = [[a,b],c]+ [b,[a,c]] \label{Jac1}
\end{equation}
holds true for all $a,b,c \in \mf g$. If we introduce the structure constants $C^a_{bc}$ by means of a basis $e_a$ of $\mf g$, $a = 1,\ldots,\dim \mf g$, \begin{equation}
[e_a,e_b]=C^c_{ab}e_c \: , \label{structure}
\end{equation} this yields the known equivalent identity
\begin{equation}\label{Jac2}
C^f_{ab}C^d_{fc} + \mathrm{cycl}_{abc} = 0~. 
\end{equation}

Lie algebras can act on a manifold $M$. This happens if one is given a Lie algebra homomorphism $\rho \colon \mf g \to \Gamma(TM)$. The vector fields in the image of $\rho$ generate a foliation. In fact, this foliation can be singular, which means that the dimension of the orbits or leaves generated in this way can have different dimensions. A typical example of this is the adjoint action where $M$ equals to $\mf g$ itself, thus $TM \sim \mf g \times \mf g$,  and $\rho(a)\vert_b = [a,b]$. Here the dimension of orbits through the generic points of $M$ equals the dimension of $\mf g$ minus its rank, while e.g.~the origin is a (zero dimensional) orbit by itself. For $\mf g = su(2)$ this foliates $M \cong \R^3$ into concentric spheres around the origin, which is the only zero-dimensional leaf in this case.

We call an \emph{almost Lie algebra} a vector space $V$ equipped with an antisymmetric product that does not necessarily satisfy the Jacobi identity \eqref{Jac1}. In some cases it may happen that the violation of the vanishing of the Jacobiator is rather mild: suppose, for example, that there exists a vector space $W$, a map $t \colon W \to V$, and an element $H \in \Lambda^3 V^* \otimes W$, which we want to interpret as a 3-bracket on $V$ with values in $W$, $H(a,b,c) := [a,b,c]$, such that 
\begin{equation}\label{3bracket}
[[a,b],c] + [[b,c],a] + [[c,a],b] =\;t\left([a,b,c]\right)
\end{equation}
holds true. In other words, the violation of \eqref{Jac1} is controlled by the image $t \circ H$ of the 3-bracket by $t$. 
Extending the 2- and 3-bracket appropriately to $V \oplus W$ (in fact, the 3-bracket is extended trivially), and requiring
additional higher Jacobi identities that contain the 3-bracket, one obtains a Lie 2-algebra (or 2-term $L_\infty$-algebra).
Suppose that the 3-bracket does not satisfy the higher Jacobi identity mentioned above, but that there exists a 
vector space $U$, a map from $U$ to $W$, and a 4-bracket governing the violation. Now there are additional Jacobi identities 
containing also the 4-bracket. If one goes on like this, one obtains the notion of a Lie $n$-algebra with $n\in \N$ or, if the
process does not stop, a Lie$_\infty$-algebra.\footnote{These are $L_\infty$-algebras concentrated in negative degrees.
The notion of an $L_\infty$-algebra goes back to \cite{Lada:1992wc}. Cf.also  for example \cite{Voronov}.} 

One of the central mathematical notions we will need in this paper is the one of a \emph{Lie algebroid}. This is a vector bundle $E$ over a manifold $M$ together with a map $\rho \colon E \to TM$, called the \emph{anchor}, equipped with a Lie algebra structure on its sections, i.e.~$(\Gamma(E),[\cdot,\cdot])$ is a Lie algebra, such that the following law, called the \emph{Leibniz rule}, holds true for all sections $s_1, s_2 \, \in \Gamma(E)$ and all $f \in C^\infty(M)$:
\begin{equation}
[s_1,f s_2] = f[s_1,s_2]+ \rho(s_1)f \cdot s_2 \, .\label{Leib}
\end{equation}
We note that if $M$ is chosen to be a point, then $E$ reduces to a vector space, sections of $E$ to elements in this vector space, the condition \eqref{Leib} becomes vacuous, and we return to the standard notion of a Lie algebra in this way. 
Next, if we are given the action of a Lie algebra $\mf g$ on a manifold $M$, this also gives rise to a Lie algebroid, called the \emph{action Lie algebroid}: one chooses $E = M \times \mf g$ and the anchor is related in an obvious way to the map determining the action, both called $\rho$ above. But by far not all Lie algebroids are action Lie algebroids. A wide class of Lie algebroids that do not originate from the action of a Lie algebra on a manifold are obtained by Poisson manifolds $(M,\Pi)$, where one chooses $E$ to coincide with $T^*M$ and the anchor is defined by means of contraction with the bivector field $\Pi$. A generalization of such examples will be provided by Dirac structures, which we are going to introduce further below.

If we choose a local basis of sections $e_a$ so as to introduce coefficients $C^a_{bc}$ as before by means of \eqref{structure} above, these coeffiients do not satisfy the condition \eqref{Jac2} in general, despite the fact that the sections form an ordinary Lie algebra. The reason is that on the one hand these coefficients now depend on the point $x \in M$, so they are genuinely \emph{structure functions} and, on the other hand, the Leibniz rule \eqref{Leib} is needed as well now to determine the bracket on general sections. Parametrising the anchor map by means of the additional structural functions $\rho_a^i$ above a  local coordinate system $x^i$ on $M$, $\rho(e_a)\vert_x=\rho_a^i(x) \partial_i$,  the Jacobi identity \eqref{Jac1} turns into
\begin{equation}
C^f_{ab}C^d_{fc} - \rho^i_a \partial_i C^d_{bc} + \textrm{cycl}_{abc}=0\: .
\end{equation}
Precisely in the case where the Lie algebroid is an action Lie algebroid, there exists a frame such that the coefficients  $C^a_{bc}$  are constant. In other words, a necessary and sufficient condition for a Lie algebroid to come from a Lie algebra action on the base is the existence of a covariantly constant frame $(e_a)$ with respect to a flat connection $\nabla$ such that $\nabla ([e_a,e_b])=0$.

We call an \emph{almost Lie algebroid} the above data, but with $(\Gamma(E),[\cdot,\cdot])$ being required to be an almost Lie algebra only. We now need to require also that $\rho$ is a morphism of brackets, which otherwise follows from \eqref{Jac1} together with \eqref{Leib}, cf., e.g., \cite{GeneralisingGeometry}. Similarly to before, some of those structures will come from a Lie $n$-algebroid (n-term $L_\infty$-algebroid) where $n \in \{ \N \cup \infty \}$. For example, for a Lie 2-algebroid, we have in addition to $E \to M$ also a vector bundle $F$ over  $M$, a fiber map $t \colon F \to E$, and a section $H \in \Gamma(\Lambda^3 E^* \otimes F)$ identified with a $C^\infty(M)$-linear 3-bracket such that \eqref{3bracket} holds true. The precise definition goes as follows: The sections in $F \oplus E$ form an (infinite-dimensional) Lie 2-algebra. In addition, there is an anchor map $\rho \colon E \to TM$ such that \eqref{Leib} holds true (for any $s_1 \in \Gamma(E)$ and $s_2 \in \Gamma(E \oplus F)$) and all the other brackets are $C^\infty(M)$-linear. Likewise for Lie n-algebroids for higher $n$.\footnote{For precise definitions of this notion in different forms we refer \cite{Bonavolonta-Poncin,Gruetzmann,Sheng-Zhu}.} 

Why are Lie algebroids and their higher $n$ generalisations of importance for us in the context of gauging? 
The reason is that any Lie $n$-algebroid generates a singular foliation on its base $M$. Suppose that $M$ is the target space of 
some scalar fields. For the context of gauge theories we are used to be given a Lie group or algebra action on $M$, which then
can be gauged. We want to permit more singular foliations to be gauged, not only those coming from group or Lie algebra actions. 
Lie algebroids provide a perfect realm for this. However, not every singular foliation can be obtained by a Lie algebroid even.
This then leads us to higher Lie algebroids: If, for example, the singular foliation is generated by an almost Lie algebroid and 
is such that the points lying in maximal dimensional leaves are dense in $M$, there always is a Lie 2-algebroid giving rise to this
foliation \cite{CamilleSylvainTS}. There are cases however, where even infinite length higher Lie algebroids are needed,
true Lie $\infty$-algebroids. In the present paper we will consider foliations coming from an almost Lie algebroid, 
be it the lower part of some Lie $n$-algebroid or not. (Some further details on these assumptions and additional information, if one has a Lie or Lie 2-algebroid, can be found in \cite{Chatzistavrakidis:2016jfz}.)

Another central notion to this paper, related to the above in a way that we are going to explain below, is the one of a Courant algebroid. Let us start with its definition. A \emph{Courant algebroid} is a vector bundle $E$ over $M$ together with an anchor map $\rho \colon E \to TM$ and a bilinear product on its sections satisfying the Leibniz rule \eqref{Leib}, as well as equation \eqref{Jac1} for all sections $a,b,c \in \Gamma(E)$. This will not be a Lie algebroid, since we will require the bracket to have a symmetric part. To control it, we introduce also a non-degenerate fiber metric $( \cdot , \cdot)$ on $E$. Together with this inner product, we can view the dual map $\rho^*$ as a map from $T^*M$ to $E$, so that for any Courant algebroid there is the sequence
\begin{equation}
0 \rightarrow T^*M \stackrel{\rho^*}{\longrightarrow} E  \stackrel{\rho}{\longrightarrow} TM \rightarrow 0 \label{exact}
\end{equation}
and the following axiom is meaningful:
\begin{equation}
[s,s] =-\frac{1}{2} \rho^* \mathrm{d}(s,s) \label{sym}
\end{equation}
to hold for any $s \in \Gamma(E)$. Having introduced a fiber metric, we also will want it to be a generalisation of ad-invariant. We thus require
\begin{equation}
2([s_1,s_2],s_2) = \rho(s_1)(s_2,s_2) \, . \label{ad}
\end{equation}
Indeed, for $M$ chosen to be a point, a Courant algebroid becomes just a quadratic Lie algebra, a Lie algebra equipped with a non-degenerate, ad-invariant scalar product. Note that in this case, and only for vanishing $\rho$, the bracket becomes automatically antisymmetric by means of \eqref{sym}. 

One might think that by anti-symmetrisation of the bracket, a Courant algebroid is also a higher Lie algebroid. This is, however,
not the case: Putting $2[s_1,s_2]_A := [s_1,s_2] - [s_2,s_1]$ it is true that this new bracket does not satisfy a Jacobi identity, 
and one might want to search for a corresponding 3-bracket to control it. But it is the Leibniz rule, valid for the 
non-symmetric bracket, which now is violated: 
$2[s_1,fs_2]_A = [s_1,f s_2] - [fs_2,s_1] = 2 [s_1,f s_2] - 2\rho^*\mathrm{d}(s_1,fs_2)$, which leads to a violation of the
Leibniz rule in form of an additional term $-2(s_1,s_2) \rho^*\mathrm{d}f$. There is, however, a way to view the algebra of 
sections of $E$ as an infinite dimensional Lie 2-algebra, with $V=\Gamma(E)$ and $W=C^\infty(M)$ \cite{Roytenberg-Weinstein}.

A particularly important class of Courant algebroids are \emph{exact} ones. By definition, they arise in the case when the sequence \eqref{exact} is exact. Then we can identify $E$ with $TM \oplus T^*M$: for any choice of splitting $\kappa \colon TM \to E$, one has $E = \kappa(TM) \oplus \rho^*(T^*M)$. Thus sections $s$ of $E$ can be identified with a couple of a vector field $v$ and a 1-form $\alpha$, $s = v \oplus \alpha$. Now one can verify that the inner product and the bracket always take the form
\begin{eqnarray}
( v \oplus \alpha,  w \oplus \beta )& =& \iota_v \beta + \iota_w \alpha \label{inner} \\
{}[v \oplus \alpha,  w \oplus \beta ]& =& {\cal L}_v w \oplus  \iota_v \mathrm{d}\beta -  {\cal L}_w \alpha - \iota_v \iota_w H \label{bracket} \, ,
\end{eqnarray}
where $H$ is a closed 3-form on $M$. Since a change of the splitting $\kappa$ leads to $v \mapsto v + \iota_v B$ for a 2-form $B$ and this changes $H$ by $\mathrm{d}B$, it is only the deRham cohomology class of $H$ which classifies different exact Courant algebroids. Since this was observed first by P.  \v Severa, $[H] \in H^3_{dR}(M)$ is called the \v Severa-class of the exact Courant algebroid. 

While the use of a Lie or a Courant algebroid $E$, so for example $TM \oplus T^*M$, as a generalisation of standard geometry,
defined for $TM$, is a philosophy that has been used in the Poisson community for quite some time, in physics it became known
in particular in the context of 2d supersymmetric sigma models \cite{Zabzine1,Zabzine2,Zabzine3} and Hitchin's paper on generalised complex 
structures \cite{Hitchin}.\footnote{Cf.~also  \cite{Gualtieri} as well as the somewhat complementary expositions \cite{HitchinINCortes} and
\cite{GeneralisingGeometry} within the same book. For a natural appearance
of the structures  \eqref{inner} and \eqref{bracket} within the context of current algebras in two dimensional models cf.\cite{AlekseevStrobl,ZabzineCurrents}.}  In the present contribution, we will report on another natural appearance of this generalised geometry in the context of two-dimensional sigma models, not in the context of supersymmetry but that of gauging. 

Consider now sections that take values inside $\rho^*T^*M \subset E$. Evidently, for such sections, the r.h.s. of \eqref{ad} vanishes and this subbundle defines a simple (in fact abelian) Lie algebroid. This can be easily deformed to something far less trivial: Consider a bivector field $\Pi$ on $M$, then its graph $\mathrm{graph}(\Pi)$, i.e. elements of the form $\iota_\alpha \Pi \oplus \alpha$, forms another subbundle of $E$ that is isotropic w.r.t. the inner product \eqref{inner} ($\rho^*T^*M \subset E$ sits as the special case $\Pi=0$ inside this class of examples). Now the bracket \eqref{bracket} does not always close for sections taking values in this subbundle; they do so precisely if the bivector satisfies the following $H$-twisted Poisson condition \cite{Park,KlimcikStrobl,Severa:2001qm} 
\begin{equation} \label{HPS}
\frac{1}{2} [\Pi,\Pi]_{SN} =\, \langle \Pi^{\otimes 3}, H\rangle \;,
\end{equation}
where the subscript $SN$ refers to the Shouten-Nijenhuis bracket and the bracket on the right denotes an appropriate contraction. Thus, any $H$-twisted Poisson bivector field $\Pi$ gives rise to a Lie algebroid structure on $D := \mathrm{graph}(\Pi) \subset TM \oplus T^*M \cong E$. This is an example of what is called a \emph{Dirac structure}: A maximally isotropic involutive subbundle of a split exact Courant algebroid. Lie algebroids are obtained in this way also, if we drop the condition of maximality: we call isotropic involutive subbundles of $E$ \emph{(regular) small Dirac structures}, while we use the specification ''\emph{irregular}'' for small Dirac structures if we talk about an isotropic involutive subsheaf of $\Gamma(E)$ without the necessity of its image to have constant rank. 

Generalised geometry now comes into the game, when one considers, for example, $E$-covariant derivatives: Let $E$ be a Lie or Courant algebroid over $M$ and $V$ another vector bundle over the same base $M$. Instead of covariant derivatives of sections $\psi$ of $V$ along vector fields $v \in \Gamma(TM)$, $\nabla_v \psi$, we now may define ${}^E\nabla_s \psi$ in a very analogous way, where $s \in \Gamma(E)$: as before the result should take values in $\Gamma(V)$, be $C^\infty(M)$-linear in the bottom argument, and now satisfy the Leibniz rule 
\begin{equation}{}^E\nabla_s (f\psi) = f {}^E\nabla_s \psi + \rho(s)f \cdot \psi \: . \label{Leib2}
\end{equation}
If then $E$ is a Lie algebroid, the standard formula for curvature gives rise to a tensor also here, and if in addition $V=E$ likewise so for torsion. An example of such $E$-covariant derivatives that will play an important role in the gauging below is induced by a normal connection $\nabla$ on $E$. We then can define  $E$-covariant derivatives on $TM$ in the following way, 
\begin{equation}
{}^E\nabla_s v := {\cal L}_{\rho(s)} v + \rho(\nabla_v s) \, , \label{Ecov}
\end{equation}
which is easily verified to be indeed $C^\infty(M)$-linear in $s$ and to satisfy \eqref{Leib2}---this is a good exercise to verify in fact. \eqref{Ecov} also induces an $E$-covariant derivative on $T^*M$ by duality, certainly. Assume now that we have \emph{two} covariant derivatives $\nabla^+$ and $\nabla^-$ on $E$, they induce two different $E$-covariant derivatives ${}^E\nabla^+$ and ${}^E\nabla^-$, and the combination
\begin{equation}
{}^E\nabla^{comb} := {}^E\nabla^+ \otimes \mathrm{id} + \mathrm{id}\otimes  {}^E\nabla^- \label{Ecovcomb}
\end{equation}
defines an $E$-covariant derivative on $T^*M \oplus T^*M$ that will play a crucial role in the sequel.

\section{On the construction of gauge theories: the traditional and the new method}

Let us now show how the above structures appear naturally in the context of gauging of $\sigma$-models. We begin with a brief reminder of the good old standard gauging of group actions. Recall that the usual procedure amounts to 
identifying some rigid symmetry of the theory, and then coupling to it gauge fields (usually minimally) 
and thus promoting the symmetry to a local one. Specialising to the case of 2D $\sigma$-models, which typically describe 
bosonic strings propagating in a target spacetime $M$, we are led to consider maps $X=(X^i): \S \to M$ from the 2D source to the 
$n$D target. The component fields $X^i$ are the dynamical fields of the theory and their action functional is given as  
\be
S_0[X]=\int_{\S}\sfrac 12 g_{ij}(X)\dd X^i\w\ast \,\dd X^j+\int_{\Sigma}\sfrac 12 B_{ij}(X)\dd X^i\w \dd X^j~.
\ee
This comprises background field data, a symmetric 2-tensor $g=(g_{ij})$ and an antisymmetric one $B=(B_{ij})$. 
Next we consider a Lie algebra $\mf g$ with elements $e_a$ mapped to vector fields $\rho_a=\rho(e_a)=\rho_a^i(X)\partial_i$ of $M$,
as explained in the previous section.
Then the action $S_0$ is invariant under the {rigid} symmetry 
$
\d_{\epsilon}X^i=\rho^i_a(X)\epsilon^a~
$
provided that:
\be
\label{standardgauge}
{\cal L}_{\rho_a}g=0~, \quad {\cal L}_{\rho_a}B=\dd\beta_a~,
\ee 
for some 1-form $\beta_a$. 

Gauging the symmetry requires the coupling  of the fields $(X^i)$ to $\mf g$-valued 1-forms $A=A^a \otimes e_a$, the \emph{gauge fields}, which is here performed by minimal coupling, i.e.~by replacing ordinary derivatives $\dd$ by  world sheet covariant ones:  
$\dd X^i \to DX^i=\dd X^i-\rho^i_a(X)A^a.$
The candidate gauged action is simply
\be \label{S12}
 S_{{m.c.}}[X,A]= \int_{\S}\sfrac 12g_{ij}(X)DX^i\w\ast DX^j+ \int_{\S}\sfrac 12B_{ij}(X)DX^{i}\w DX^{j}~.
\ee
The action is invariant under the (standard) infinitesimal gauge transformations:
\bea
\d_{\epsilon}X^i&=&\rho^i_a(X)\epsilon^a~,\\
\d_{\epsilon}A^a&=&\dd \epsilon^a+C^a_{bc}A^b\epsilon^c~,
\eea
with  a $\Sigma$-dependent gauge parameter $\epsilon^a$ provided that  $\beta_a=0$.
Note that for $\beta_a\ne 0$, minimal coupling is not sufficient \cite{Hull:1989jk}; this is discussed later on. 

{Our purpose is to go beyond this standard gauging procedure.}
Thus we default and withdraw the requirement for a rigid symmetry; as a result there are no specified initial assumptions 
for $g(X)$ and $B(X)$.
In other words, considering again the candidate (minimally-coupled) \emph{gauged} action,
we ask the following question:
\emph{Under which conditions does $S_{{m.c.}}$ have a gauge 
symmetry $\d_{\epsilon}X^i=\rho^i_a(X)\epsilon^a$ for \emph{some} accompanying transformation formulas for the gauge fields $A^a$?}
In order to answer this question, first we generalize our assumptions in the direction of Section 2. In particular, we
replace the Lie algebra $\mf g$ by \emph{some} vector bundle $E \overset{\pi}
\to M$ with an \emph{almost Lie} algebroid structure
$$
E
\overset{\rho}
\to TM~,\quad  [\cdot,\cdot]_E\, .$$
In a local basis of sections $e_a$ of $E$:
\begin{equation}\label{Cabc}
[e_a,e_b]_{E}=C_{ab}^c({x})e_c \quad 
\to \quad [\rho_a,\rho_b]_{{Lie}}=C_{ab}^c(x)\rho_c~.
\end{equation}
Thus the vector fields $\rho_a=\rho(e_a)$ are involutive and generate a
 (possibly singular) foliation ${\cal F}$ on $M$.  Let us now make the most general Ansatz for the gauge transformation of $A^a$:
\bea 
\d_{\epsilon} A^a&=&\dd \epsilon^a+C^a_{bc}(X)A^b\epsilon^c{+\Delta A^a}~,
\eea
for some, at this point unspecified contributions $\Delta A^a$.
The transformation of \eqref{S12} reads as 
\begin{eqnarray}
\delta_\epsilon S_{{m.c.}}&=& \int_{\S}\epsilon^a\biggl(  \sfrac 12({\cal L}_{\rho_a}g)_{ij}D X^i\w\ast D X^j+\sfrac 12 ({\cal L}_{\rho_a}B)_{ij}D X^{i}\w  D X^{j}\biggl) \nonumber \\
&&\!\!\!\!\!\!\!\!\! - \int_{\S} g_{ij} \rho_a^i \Delta A^a \wedge  \ast D X^j + B_{ij}\rho_a^{i} \Delta A^a \wedge D X^{j}\, . 
\end{eqnarray} 
Considering{\footnote{This refinement of the Ansatz is not the most general one. For a more complete discussion cf.\cite{Chatzistavrakidis:2016jfz}.}} 
${\Delta A^a = \omega^a_{bi}(X)\epsilon^bDX^i + \phi^a_{bi}(X)\epsilon^b \ast DX^i,}$
 invariance of $S_{{m.c.}}$ requires
{}{{\bea 
\label{condition1gB} {\cal L}_{{\rho}_a}g&=&\o^b_a\vee\iota_{{\rho}_b}g-\phi^b_a\vee \iota_{{\rho}_b}B~, 
\\
\label{condition2gB} {\cal L}_{{\rho}_a}B&=&\o^b_a\w\iota_{{\rho}_b}B\pm \phi^b_a\w \iota_{{\rho}_b}g~,
\eea}}
where the undertermined sign refers to the signature (Euclidean/Lorentzian) of the worldsheet and is controlled by the square of the Hodge operator: 
$\ast^2=\mp 1$.

In order to understand the geometric interpretation, we examine 
what happens under a change of basis $e_a \to \L^b_ae_b$ in $E$:
\bea\label{transo}
\o^a_{bi}\to (\Lambda^{-1})^a_c\o^c_{di}\Lambda^d_b-\Lambda^c_b\partial_i(\Lambda^{-1})^a_c~,\qquad 
\phi^a_{bi}\to (\Lambda^{-1})^a_c\phi^c_{di}\Lambda^d_b~. \eea
These transformation rules clearly indicate that  $\o^a_{bi}$ are the coefficients of a connection 1-form on the vector bundle $E$:
\be 
\nabla^{\omega} e_a=\omega_a^b\otimes e_b~,
\ee
and $\phi^a_{bi}$ are the coefficients of an endomorphism-valued 1-form: $\phi\in\G(T^{\ast}M\otimes E^{\ast}\otimes E)$.
Moreover, since the difference of two vector bundle connections is an endomorphism 1-form,
this means that the gauging is controlled by two connections $\nabla^{\pm}=\nabla^{\o}\pm\phi$ on $E$. 

On the other hand we observe a mixing of $g$ and $B$, which is reminiscent of T-duality and generalized geometry.
One way to explain this uses the combination \eqref{Ecovcomb} of $\nabla^+$ and $\nabla^-$ introduced in the previous section.
In \cite{Kotov:2016lpx} it was shown that Eqs. \eqref{condition1gB} and \eqref{condition2gB} can be combined into one elegant formula:
\begin{equation} {}^E\nabla^{comb} (g+B)= 0~.\label{Alexeiformula}\end{equation}
A slightly different viewpoint in terms of graphs may be found in \cite{Chatzistavrakidis:2016jci}.

So far we have discussed what happens when the gauge fields are introduced in the theory through minimal coupling.
However, in the presence of a Wess-Zumino term in the action, or already  when $\beta_a$ in \eqref{standardgauge} does not vanish, 
minimal coupling is not enough. Starting with the ungauged action
\begin{equation} \label{S0WZ}
S_{0,WZ}[X] = \int_{\S}\sfrac 12 g_{ij}(X)\dd X^i\w\ast \,\dd X^j + \int_{\hat \S} \sfrac 1{3!}H_{ijk}\dd X^i\w\dd X^j\w\dd X^k \, ,
\end{equation}
the candidate gauged action, at least for minimally coupled kinetic sector, is
\begin{equation} \label{ansatz1}
S_{WZ}[X,A]=\int_\Sigma\sfrac{1}{2} g_{ij}(X) D X^i \w \ast D X^j + \int_{\hat\S} H + \int_\Sigma A^a \wedge \theta_a +
\sfrac{1}{2} \gamma_{ab}(X) A^a \wedge A^b \: , 
\end{equation}
where $\theta_a=\theta_{ai}(X)\dd X^i$ are 1-forms and $\gamma_{ab}(X)$ are functions on $M$, both pulled back via $X$.
Such an action was first studied in \cite{Hull:1989jk} and more recently in \cite{Plauschinn:2013wta,Kotov:2014dqa,Bakas:2016nxt}.
As in the minimally coupled case, we examine the gauge invariance of $S_{WZ}$ under the transformations $\d_{\epsilon}X^i$ and 
$\d_{\epsilon}A^a$.
 The new invariance conditions are found to be 
 {\bea 
\label{condition1gH} {\cal L}_{{\rho}_a}g&=&\o^b_a\vee\iota_{{\rho}_b}g+\phi^b_a\vee \theta_b~,
\\
\label{condition2gH} \iota_{\rho_a} H &=&\dd \theta_a - \o^b_a\w\theta_b \pm \phi^b_a\w \iota_{{\rho}_b}g~,
\eea }
to hold true for \emph{some} choice of $\nabla^{\pm}$ (encoded in terms of $\o^b_a$ and $\phi^b_a$ as before) as well as of the 1-forms $\theta_a$ on $M$. 
There are two additional obstructing constraints
{\begin{equation}\label{Calpha}\iota_{\rho_a} \theta_b +  
\iota_{\rho_b} \theta_a = 0~, \quad 
\iota_{\rho_b}\iota_{\rho_a}H=C^d_{ab}\theta_d+\dd\iota_{\rho_{[a}}\theta_{b]} -2{\cal L}_{\rho_{[a}}\theta_{b]}~.
\end{equation} }
The quantities $\gamma_{ab}$ in \eqref{ansatz1} are then uniquely determined from these data: $\gamma_{ab}= \iota_{\rho_a} \theta_b$.

In the subsequent section, we will interpret the conditions \eqref{Calpha} geometrically, thus relating also the previous two sections in the context of generalised geometry.

\section{Universal action functional and its generalised geometry}

In this section we present our main result: the existence of a universal action functional that describes all the gaugings from before. We will go step by step and first define the functional, given for the same data as the ungauged theory \eqref{S0WZ} for which it will describe all possible gaugings, and only then relate it to the previous two sections, marrying them into a beautiful picture of gauge theories and generalised geometry underlying it. 

The action functional  \eqref{S0WZ} is defined for (essentially) maps $X \colon \Sigma \to M$\footnote{We drop subtleties about Wess-Zumino terms here, for which we need to extend the map $X$ appropriately; the statements remain correct, if considered as a variational problem.} Let us rewrite it into an, at this point, absolutely equivalent theory, defined now with two additional auxiliary 1-form fields, $V^i$ and $W_i$. These two fields are 1-forms on $\Sigma$ with values in $T^*M$ and $TM$, respectively; together they combine into  $V\oplus W \in \Omega^1(\Sigma, X^*TM \oplus X^*T^*M)$. Thus, $X$ and the 1-form fields correspond precisely to a vector bundle morphism $u \colon T\Sigma \to TM \oplus T^*M$, i.e.~the source of the sigma model is extended to the \emph{tangent} bundle of the worldsheet $\Sigma$ and the target is to the generalised tangent bundle of the previous target $M$. 

We now define the action functional that will play a fundamental role in all what follows: 
\begin{equation}\label{univ}
  S_{univ}[u] =\, \int_\Sigma \, \sfrac 12 g_{ij}\,{\cal D} X^i \wedge \star {\cal D}X^j + \int_{\hat \Sigma}\,H + \int_\Sigma \, W_i \wedge (\dd X^i - \tfrac{1}{2} V^i)\;,
\end{equation}
with ${\cal D}X^i = \dd X^i - V^i$, and where we assume $\hat \Sigma$ to be a three-dimensional manifold with boundary $\Sigma$ (and pretend that this extension depends only on $u$ for simplicity, even if this is not completely true, cf the previous footnote). 

We claimed above that this is equivalent to the functional  \eqref{S0WZ}, and in fact so classically as well as quantum mechanically. Let us demonstrate this explicitly first: Variation with respect to the auxiliary fields $V$ and $W$ permit to express these two fields uniquely and in an algebraic fashion:
\begin{equation}
V = 2 \dd X \quad, \qquad W = 2 \ast g( {\cal D}X, \cdot  )
\end{equation} 
In such a situation, one is permitted to implement the expressions for the pair of fields $V$ and $W$ back into the action. Doing so in \eqref{univ}, we see that the last term vanishes identically and, noting that ${\cal D}X^i = - \dd X^i$, we see that indeed this leads directly back to \eqref{S0WZ}. 

Now one may ask certainly for what this reformulation is good then, if it is just an equivalent formulation of the original functional. The point is the following one: The theory \eqref{univ} \emph{changes} drastically for the case that the fields $V$ and $W$ are constrained, so as to take values in a subspace inside $TM \oplus T^*M$. But, most importantly, \emph{every} gauging of this model, be it a standard gauging of a group $G$ acting on $M$ or the gauging of a foliation, is of this form! We will now explain this in detail in what follows. 

To establish a relation between the action \eqref{ansatz1} and the above one, \eqref{univ}, we observe that already the kinetic term fixes the relation between $A$ and $V$ and the remaining part of the actions linear in $\dd X$ the one between $A$ and $W$: \begin{equation}
    \label{VWdef}
    V^i =\, X^*(\rho^i_a) A^a\;, \quad W_i =\, X^*(\theta_{ai}) A^a\; .
  \end{equation}
At this point now it is absolutely crucial that gauge invariance of the action \eqref{ansatz1} requires $\gamma_{ab} = \,\rho^i_a \theta_{bi}$. It is precisely this expression that is needed (up to an anti-symmetrisation, which anyway is dictated/ensured by the first equation in \eqref{Calpha}) to make the remaining terms of the two actions agree, nothing else would do! 

Let us now recall that $A$ (together with the original field $X$) corresponds to a vector bundle morphism from $T\Sigma$ to the almost Lie algebroid $E$, which in turn is equipped with its anchor map $\rho \colon E \to TM$. Note that the first relation in \eqref{VWdef} contains precisely this anchor and thus is implicitly determined by the choice of the foliation we want to factor out. In other words, the information fixed by the first equation in \eqref{VWdef} fixes the foliation to be gauged out in the ungauged theory \eqref{S0WZ} or \eqref{univ}.  What is the information contained in the second equation \eqref{VWdef}? Evidently it corresponds to lifting the map $\rho$, specified in the first equation, to a map $\sigma \colon E \to TM \oplus T^*M$, as illustrated in the following (commutative) diagram.
 \begin{equation}\label{FET*M}
 \xymatrix{&  \:\,TM\oplus T^*M \ar[d]
  \\
     E \ar[r]^{\mathlarger{\;\,\rho}} \ar@{-->}[ru]^{\mathlarger{\sigma}}  & TM}
  \end{equation}
  These data now permit to  relate the ungauged action \eqref{univ} with its auxiliary fields $V\oplus W$ to the general gauged action \eqref{ansatz1} found in the previous section. More precisely, the latter functional is defined on vector bundle morphisms $u \colon T\Sigma \to TM \oplus T^*M$; we denote the set of such vector bundle morphisms $u$ by ${\cal U}$. Similarly, we denote the set of vector bundle morphisms $a \colon  T\Sigma \to E$, corresponding to the  couple $(X,A)$ appearing in \eqref{ansatz1},  by ${\cal A}$. Now, evidently, for any  choice of $\sigma$ in \eqref{FET*M}, we can construct a map 
  \begin{equation}
  \widehat{\sigma} \colon {\cal A} \to {\cal U} \; , \quad a \mapsto \sigma \circ a
  \end{equation}
such that the following commutative diagram holds true.
\begin{equation}\label{diagrammain}
  \xymatrix{ & \:\, \circled{{\cal U}} \ar[d]^{{S_{univ}}}
  \\
    \circled{{\cal A}} \ar@{-->}[ur]^{\mathlarger{\exists ! \:\,\widehat\sigma}}   \ar[r]^{{\:\,\, {S_{{gauged}}}}} & \R} 
\end{equation}
In other words, $S_{gauged}[a]$ as in \eqref{ansatz1} can be obtained from \eqref{univ} together with \eqref{FET*M} in the form
  \begin{equation} \label{pullback}
 S_{gauged}[a]  = S_{univ}[\sigma \circ a] \qquad \Leftrightarrow \qquad  S_{gauged} =  \widehat{\sigma}^* S_{univ} \, .
  \end{equation}
The exclamation mark in \eqref{diagrammain} signals that the map $\widehat{\sigma}$ is unique for a given choice of $S_{gauged}$ for the diagram or, equivalently, \eqref{pullback} to hold true. This is an essential property for what mathematicians call a ''universality property''. That it is unique follows directly from the uniqueness that we found in \eqref{VWdef}.

Up to now, we dealt with the generalised tangent bundle $TM \oplus T^*M$, but in this section no where did we use any of its structure relating to the exact Courant, \eqref{inner} and \eqref{bracket}. In fact, they enter, if we reverse the question from above. We asked here, \emph{given} $S_{gauged}$, can it always be obtained in a unique way from a functional that exists irrespective of the foliation that we gauged out in $S_{gauged}$ and the answer to this question was affirmative. Let us turn the question around now: we start with the ungauged theory. We learned above that for this we can equally consider the functional \eqref{univ}. We now want to gauge some foliation given by some map $\rho \colon E \to TM$. How do we need to proceed? Gauging in the above, new picture, does not mean \emph{extending} a given functional \eqref{S0WZ}, it now means instead \emph{restricting} or constraining the fields in the equivalent ungauged theory \eqref{univ}. We learn from the above picture that now gauging a string theory with target $(M,g,H)$ is equivalent to choosing the data in \eqref{FET*M}. But what are admissible choices for the map $\sigma$?

It is here, where the generalised geometry comes into the game. The constraints to be posed on the choice of data, after fixing the foliation generated by $\rho$, i.e.~by $\rho_a = \rho(e_a)$ in some arbitrarily chosen local basis of sections of $E$, we need to choose $\theta_a = \theta(e_a)\equiv \theta_{ai}\dd x^i$,  such that the constraints \eqref{Calpha} hold true. Here $\theta \colon E \to T^*M$ is the missing part of the lift $\sigma = \rho \oplus \theta$ in \eqref{FET*M}. We now specify the structural formulas \eqref{inner} and \eqref{bracket}, defining an $H$-twisted standard Courant algebroid, as one may also call a split exact Courant algebroid for the case that $H$ is the representative of its Severa class for a given splitting $\kappa$ of the sequence \eqref{exact}, when restrained to the image of $\sigma$: 
\begin{equation} \label{Diracconditions} (   \sigma(e_a),\sigma(e_b) )= \iota_{\rho_a}\theta_b+\iota_{\rho_b}\theta_a \; ,\qquad [\sigma(e_a),\sigma(e_b)] = \,C^c_{ab} \rho_c \oplus {\cal L}_{\rho_a} \theta_b -\iota_{\rho_b} d\theta_a - \iota_{\rho_a}\iota_{\rho_b} H
\end{equation}{}where in the second expression we made use of \eqref{Cabc}. 
 Comparing these equations with \eqref{Calpha}, we see that the latter constraints are \emph{equivalent} to 
\begin{equation}   (   \sigma(\Gamma(E)),\sigma(\Gamma(E)) ) = 0 \; ,\qquad   [\sigma(\Gamma(E)),\sigma(\Gamma(E))]   \subset \sigma(\Gamma(E)) \, .
\end{equation}
In other words, $\sigma$ in \eqref{FET*M} has to be chosen such that \emph{its image, i.e.~$\sigma(\Gamma(E))$, forms a (possibly irregular small) Dirac structure}.\footnote{Cf.~section 2 for the nomenclature.}

After this has been accomplished, one needs to still  solve for the conditions \eqref{condition1gH} and  \eqref{condition2gH}. These two equations need to hold for the two sought-for connections $\nabla^\pm$, parametrised by the local 1-forms $\omega^a_b$ and $\phi^a_b$. In principle they may or may not have solutions, depending on the choice of the small Dirac structure. Observe, that it is only here where the metric $g$ as well as the two connections $\nabla^+$ and $\nabla^-$ come into the game. It is thus very suggestive to discuss those constraints in a remaining step. They can be viewed of generalising the invariance conditions within the standard setting on the target data, like  \eqref{standardgauge} for the case of $g$ and $B$. Although it is not difficult to express the equations  \eqref{condition1gH} and  \eqref{condition2gH} in some basis-independent manner, cf.~\cite{Chatzistavrakidis:2016jfz}, an equally nice reformulation of them such \eqref{Alexeiformula} for $g$ and $B$ is still missing as of the moment.

The universality property of the action $S_{univ}$, the form of which was  found in the context of topological models already in \cite{Kotov:2004wz}, was remarked first in \cite{Kotov:2014dqa}; it was  proven there for the standard gauging of group symmetries and announced for the gauging of foliations, proven rigorously in \cite{Chatzistavrakidis:2016jfz}. The relation of gauging in 2d and Dirac structures were noted first in \cite{AlekseevStrobl} within a Hamiltonian framework. The findings explained in this section represent its Lagrangian counterpart.

\newpage
\section{The example of the $H$-twisted Poisson Sigma Model with kinetic term}

In this section, we discuss a relatively simple topological theory for illustrative reasons, namely the Poisson $\sigma$-model \cite{Ikeda:1993fh,Schaller:1994es}
\be 
S_{{PSM}}[X,A]=\int_{\S}A_i\w\dd X^i+\sfrac 12 \Pi^{ij}(X)A_i\w A_j~. \nn
\ee
This model has several attractive features, among which are that  its path integral quantisation yields Kontsevich $\star$ product \cite{Cattaneo:1999fm} and that it can be seen to host a large class of two-dimensional Yang-Mills theories coupled dynamically to various 2d gravity models, with and without torsion \cite{KloeschStrobl}. Adding a 3-form WZ-term $H$ to this model, one obtains the $H$-twisted Poisson sigma model \cite{KlimcikStrobl}; here $\Pi$ has to satisfy the twisted Poisson condition \eqref{HPS}. In \cite{Kotov:2004wz} it was shown that the topological character of the $H$-twisted Poisson sigma model remains completely unchanged when adding a kinetic term to this action: 
\begin{equation}\label{PSMg}
S_{gHPSM}[X,A] = \int_{\Sigma} \,\left( A_i \wedge \dd X^i + \sfrac{1}{2}\Pi^{ij}A_i \wedge A_j\right)+ \int_{\hat\Sigma} \, H
+ \; \int_{\Sigma}\, 
\sfrac{1}{2}g_{ij}(X)D X^i\w\ast D X^j ~,
\end{equation}
where $DX \equiv \dd X - \Pi(A, \cdot)$. 
In fact, for an \emph{arbitrary} choice of the metric $g$ on $M$, the gauge symmetries of the model without kinetic term get deformed only, but never disappear.\footnote{It was these kind of observations, the existence of a gauge symmetry of \eqref{PSMg} independently of the fact that $g$ has Killing vectors or not, that led to \cite{withoutsymmetry} and all the present considerations ultimately.}

This implies, read backwards, that the addition of the first part in \eqref{PSMg} together with the replacement $\dd X \mapsto DX$, consists of a possible gauging of \eqref{S0WZ} for the foliation that is induced by the bivector field $\Pi$. We now show in detail how to obtain this model of a gauging when starting from the ungauged theory in the form $S_{univ}$ by restriction: The Dirac structure we choose is the one of the graph of $\Pi$, $D=\mathrm{graph}(\Pi) \equiv \{ \iota_\alpha \Pi \oplus \alpha , \alpha \in T^*M \}$. 
Restricting the auxiliary fields $V \oplus W$ in  \eqref{univ} to take values in the Dirac structure, i.e.~to be elements of $\Omega^1(\Sigma X^*D)$, we can parametrize them  by an unconstrained gauge field $A\in\Omega^1(\Sigma,X^*T^*M)$  as follows  $$W_i=A_i \quad {\rm and} \quad 
V^i=\Pi^{ji}A_j~.$$ Plugging this into \eqref{univ}, we indeed obtain \eqref{PSMg}.

It remains to solve the constraints \eqref{condition1gH} and \eqref{condition2gH}. One can verify that this can be achieved in the above holonomic basis by means of the following choice, valid (and well-defined) for arbitrary choice of $g$: 
\bea\label{h0}
 \o^j_{ik}&=&
 \Gamma^j_{ik}+g_{il}\Pi^{lm}\phi^j_{mk}+\sfrac 12\Pi^{jl}H_{lik}~,
 \\
 \phi^j_{ik}&=&
 -[(1-g\Pi g\Pi)^{-1}]^l_ig_{lm}(\mathring\nabla_k\Pi^{mj}+\sfrac 12 H_{knp}\Pi^{nm}\Pi^{pj}) 
 ~,\nn\eea
 where $\mathring\nabla$ is the Levi-Civita connection of $g$.

It is remarkable that constraining the auxiliary 1-form fields to take values in a Dirac structure, they change their nature so drastically so as to become gauge fields. The constrained theory has a huge gauge symmetry of the form 
discussed in section 3. For a better comparison, we remark that here we can identify $E$ with $T^*M$ and equip it with the  $H$-twisted Koszul-Schouten bracket between 1-forms $\a,\widetilde{\a}$:
\bea \label{kosH}
[\a,\widetilde\a]_{{KS}}:={\cal L}_{\iota_\a \Pi}\widetilde\a-\iota_{\iota_{\widetilde\a} \Pi}\dd \a
-H(\iota_{\widetilde\a} \P,\iota_{\a} \P,\cdot)~,\nn
\eea
which in a holonomic basis ${\dd x^i}$ of local sections of $T^{\ast}M$ satisfies $[\dd x^i,\dd x^j]_{\rm KS}=C^{ij}_k(X)\dd x^k$ 
with the structure functions $C^{ij}_k=\partial_k\Pi^{ij}+ H_{kmn}\Pi^{mi}\Pi^{nj}$. It is these coefficient functions that enter the gauge transformations together with the coefficients \eqref{h0}.
 
We finally remark that the gauging is possible for \emph{every} full Dirac structure. In \cite{Chatzistavrakidis:2016jci} we provided the generalisation of \eqref{h0} to arbitrary Dirac structures in a closed form. This provides an interpretation of the gauge symmetries of the Dirac sigma model \cite{Kotov:2004wz} in the spirit of the gauging of foliations advocated here.

 \section{Non-topological examples of the movement of strings in quotient spaces}
 
 As a simple class of non-topological examples, we examine a 4D nilmanifold and a host of possible quotients. Recall that nilmanifolds are compact homogenous spaces $G/H$, where $G$ is a (compact) nilpotent Lie group and $H$ a discrete, cocompact subgroup of $G$. There exist three nilmanifolds in 4D, one being simply the 4-torus, 
 and two being toroidal fibrations of different step; in particular, there is a step-2 4D nilmanifold,\footnote{A nilpotent Lie algebra is of step-$i$, if its lower central series terminates after $i$ 
 steps, i.e.~if one has $\mf g_{\{i\}} \equiv [\mf g,\mf g_{\{i-1\}}]=0$, where $\mf g_{\{0\}}\equiv \mf g$. Thus abelian Lie algebras are step-1 and the corresponding nilmanifolds are simply tori.}
 essentially the 3D Heisenberg
 nilmanifold{\footnote{We refer to \cite{Chatzistavrakidis:2015lga,Chatzistavrakidis:2016vsy} for a discussion of gauging and T-duality for such step-2 cases.}} 
 trivially fibered over a circle, and a step-3 one, which we examine here. All three cases admit a symplectic structure too. 
 
 The case we examine here is based on the following step-3 nilpotent Lie algebra with four generators $\{e_a\}, a=1,2,3,4$:
 \be 
 [e_1,e_2]=e_4~, \quad [e_2,e_4]=-e_3~. 
 \ee
 In other words, we take structure constants $C_{12}^4=C_{42}^3=1$, all the remaining ones vanishing. The corresponding nilmanifold is the quotient of the compact Lie group of this algebra by the left action $\rho$ of a given discrete subgroup. The resulting smooth geometry can be equipped with the symplectic structure 
  \be 
  \Omega=e^1\w e^4+e^2\w e^3~,
  \ee
where $e^a$ are 1-forms dual to the vector fields $\rho_a=\rho(e_a)$. We hereby choose coordinates where they take the explicit form
\be 
e^1=\dd x^1~, \quad e^2=\dd x^2~, \quad e^3=\dd x^3-(x^4+x^1x^2)\dd x^2~, \quad e^4=\dd x^4+x^2\dd x^1~.
\ee
We then observe that $\Omega=\dd x^1\w\dd x^4+\dd x^2\w\dd x^3$. The dual vector fields are given as 
\be \label{vfs}
\rho_1=\partial_1-x^2\partial_4~, \quad \rho_2=\partial_2+(x^4+x^1x^2)\partial_3~, \quad \rho_3=\partial_3~, \quad \rho_4=\partial_4~,
\ee
and they are globally well-defined. 
A natural line element on this manifold is given as $\dd s^2=\d_{ab}e^ae^b$. The corresponding ungauged $\sigma$-model, with vanishing $B$-field, is 
\bea 
S[X]&=&\int_{\S}\sfrac 12 \left(1+(X^2)^2\right)\dd X^1\w\ast\dd X^1+\sfrac 12 \left(1+(X^4+X^1X^2)^2\right)\dd X^2\w\ast\dd X^2+ 
\sfrac 12 \dd X^3\w\ast\dd X^3 + \nn\\ && +\int_{\S}\sfrac 12 \dd X^4\w\ast\dd X^4-(X^4+X^1X^2)\dd X^2\w\ast \dd X^3 +
X^2\dd X^1\w\ast\dd X^4~.
\eea
Now we would like to gauge this model along some vector fields. For simplicity, here we follow mainly the more standard extension approach along the lines of section 3. 

A pool of options is given by the vector fields that 
appear in Eq. \eqref{vfs}. First we note that only $\rho_3$ is a Killing vector for the metric. For the other three vector fields we find
\bea \label{Lieg}
{\cal L}_{\rho_1}g&=&-x^2\dd x^1\vee \dd x^2-\dd x^2\vee \dd x^4~, \nn\\
{\cal L}_{\rho_2}g&=& x^2\dd x^1\vee\dd x^1 -x^2(x^4+x^1x^2)\dd x^1\vee\dd x^2+x^2\dd x^1\vee\dd x^3+\dd x^1\vee\dd x^4-
 \nn\\ &&-(x^4+x^1x^2)\dd x^2\vee\dd x^4+\dd x^3\vee\dd x^4~,\nn\\
{\cal L}_{\rho_4}g&=&(x^4+x^1x^2)\dd x^2\vee\dd x^2-\dd x^2\vee\dd x^3~.
\eea

We have shown in the preceding section that if we take $E=D$, i.e., if we gauge along all four vector fields forming a
full Dirac stucture, we can find the connection coefficients $\omega_a^b$. A solution{\footnote{In general there is a multi-parameter family of 
solutions; here we present one regular such solution.}} of Eq. \eqref{condition1gB} is:
\bea
&&\o_1^1=-x^2\dd x^2~,\quad \o_1^2=-\dd x^4~, \quad \o_2^1=- x^2(x^4+x^1x^2)\dd x^2~,\nn\\ && \o_2^2=-(x^4+x^1x^2)\dd x^4, \quad \o_2^4=\dd x^1+\dd x^3~, \quad \omega_4^3=-\dd x^2~.
\eea
However, in this case one would end up gauging away all four dimensions resulting into a  purely topological model again, which is not the goal of the present section.
For a non-topological gauge theory we need to choose $E$ to be a \emph{small} Dirac structure. 
Which are the one-, two-, or three-dimensional foliations we can gauge for the 4D manifold at hand? 

Clearly, we can gauge along the Killing vector $\rho_3$  generating a regular foliation. The gauged action is 
\bea 
S[X,A^3]&=&\int_{\S}\sfrac 12 \left(1+(X^2)^2\right)\dd X^1\w\ast\dd X^1+\sfrac 12 \left(1+(X^4+X^1X^2)^2\right)\dd X^2\w\ast\dd X^2+ 
\sfrac 12 DX^3\w\ast DX^3 + \nn\\ && +\int_{\S}\sfrac 12 \dd X^4\w\ast\dd X^4-(X^4+X^1X^2)\dd X^2\w\ast DX^3 +
X^2\dd X^1\w\ast\dd X^4~,
\eea
where $DX^3=\dd X^3-A^3$. Variation of the action w.r.t.$A^3$ gives $DX^3=(X^4+X^1X^2)\dd X^2$ or $A^3=e^3$. 
The equations of motion for $X^i$ (using the gauge constraint) are all regular:
\bea
&& \delta_{X^1}S=0 \, \Rightarrow \dd\ast\left((1+(X^2)^2)\dd X^1+X^2\dd X^4\right)=0~, \\ 
&& \delta_{X^2}S=0 \, \Rightarrow \dd\ast\dd X^2-\dd X^1\w\ast(\dd X^4+X^2\dd X^1)=0~, \\
&& \delta_{X^4}S=0 \, \Rightarrow \dd\ast(\dd X^4+X^2\dd X^1)=0~.
\eea
 Integrating out the gauge field, which enters the action quadratically, we obtain
\bea 
S_{\rm red}[X]&=&\int_{\S}\sfrac 12 \dd X^1\w\ast\dd X^1+\sfrac 12 \dd X^2\w\ast\dd X^2+ \sfrac 12 (\dd X^4+X^2\dd X^1)\w\ast (\dd X^4+X^2\dd X^1)~,
\eea
a reduced worldsheet action for the 3D Heisenberg nilmanifold. This exhausts the isometric possibilities. Let us now look for 
non-isometric ones. 

The first set of options is to attempt a gauging along any single one of the global vector fields $\rho_1,\rho_2,$ or $\rho_4$. 
Each of them can be rather trivially seen as a small Dirac structure. However, using the gauged actions obtained by minimal coupling, 
none of these vector fields or foliations leads to a consistent gauge theory, in the 
sense that no corresponding solution can be found for $\omega_a^b$ in the corresponding equation \eqref{condition1gB} and \eqref{condition2gB} (in fact, for vanishing $B$, the second equation becomes vacuous). 

The second set of options comprises small Dirac structures generated by two vector fields.
These are given by the following four pairs of commuting vectors{\footnote{The other pairs of vector fields do not constitute small Dirac structures because they are not 
involutive.}}: $(\rho_1,\rho_3),$  $(\rho_1,\rho_4),$  $(\rho_2,\rho_3)$ and $(\rho_3,\rho_4)$. 
One can show that only the small Dirac structure generated by $(\rho_3,\rho_4)$  provides a consistent gauging for the worldsheet 
action in the above sense, while for the rest of them there is no $\omega_a^b$ such that Eq. \eqref{condition1gB} can be solved. For the solvable case, 
the simplest choice of $\omega_a^b$ is just
\be 
\omega_4^3=-\dd X^2~,
\ee
with all other components vanishing. The string then propagates in the quotient, which is a 2-torus:
\be 
S[X^1,X^2]=\int_{\S}\sfrac 12 \dd X^1\w\ast\dd X^1+\sfrac 12 \dd X^2\w\ast\dd X^2~.
\ee

Finally, the third and last set of options regards small Dirac structures generated by three vector fields. 
These are two---one given by the set of commuting vectors $(\rho_1,\rho_3,\rho_4)$ and the other defining the subalgebra
$[\rho_2,\rho_4]=-\rho_3$. Both of these small Dirac structures can be used to produce a consistent gauging of the original 
worldsheet action, leading to two different quotients. In the toroidal case of $(\rho_1,\rho_3,\rho_4)$, we solve the invariance 
condition \eqref{condition1gB} with 
\be 
\omega_1^4=-\dd X^2~, \quad \omega_4^3=-\dd X^2~.
\ee
The resulting quotient is simply the $X^2$-circle. Similarly, when the vector fields are chosen to be  $(\rho_2,\rho_3,\rho_4)$, 
the invariance condition is solved by
\be 
\omega_2^4=\dd X^1-(X^4+X^1X^2)\dd X^2+\dd X^3~, \quad \omega_4^3=-\dd X^2~.
\ee
The quotient then is an $X^1$-circle. 
We summarise the above findings in Table \ref{tab1}.

\begin{table}\centering
\begin{tabular}{cccc}
 {Small Dirac Structure $D_{(\rho_a)}$} & {Leaves} & {Existence of Gauging} & {Target Space}
 \\[4pt]\hline \\
 $D_{(\rho_1)}$ & circles & no & n/a 
 \\[4pt]
 $D_{(\rho_2)}$ & circles & no & n/a
 \\[4pt]
 $D_{(\rho_4)}$ & circles & no & n/a
 \\[4pt]
 $D_{(\rho_1,\rho_3)}$ & 2-tori & no & n/a \\[4pt]
 $D_{(\rho_1,\rho_4)}$ & 2-tori & no &  n/a \\[4pt]
 $D_{(\rho_2,\rho_3)}$ & 2-tori & no & n/a \\[4pt]
 $D_{(\rho_3,\rho_4)}$ & 2-tori & yes & 2-torus \\[4pt]
 $D_{(\rho_1,\rho_3,\rho_4)}$ & 3-tori & yes & circle \\[4pt]
 $D_{(\rho_2,\rho_3,\rho_4)}$ & 3-nilmanifolds & yes & circle
 \end{tabular}
 \caption{Small Dirac structures and options for gauging foliations of the 4D step-3 nilmanifold generated by non-Killing 
 vector fields $\rho_a$ using minimal coupling. There still may be other solutions, in particular, if we do not restrict to non-minimal coupling, but use the universal action instead. This happens, for example, in the case of  the foliation generated by $\rho_4$ in the case of Euclidean worldsheets, while  the $\rho_1$-foliation  does not permit a gauging also within the enlarged setting.}
 \label{tab1}
 \end{table}

There may be more options for gauging non-isometry directions in the case that we do not restrict to minimal coupling, but that we use instead the more general functional \eqref{ansatz1}. Its invariance conditions, given by equations \eqref{condition1gH} and \eqref{condition2gH}, now have further parameters to tune a solution: In addition to $\omega^a_b$, we now also have $\theta_a$ and $\phiâ_b$ at our free disposal. We do not perform an equally systematic study of this enlarged framework for the above system, but provide the result for two examples: The foliation generated by $\rho_1$ still cannot be gauged in this way. However, in the case of $\rho_4$, there now is a Dirac structure, different from the trivial $D_{(\rho_4)}=\{ f \rho_4 \oplus 0 , f \in C^\infty(M) \}$, that does permit gauging. One verifies that $D'_{(\rho_4)}=\{ f [\rho_4 \oplus  \left((x^1x^2+x^4)\dd x^2 - \dd x^3\right)], f \in C^\infty(M) \}$ is a small Dirac structure, i.e.~its elements satisfy the two conditions \eqref{Diracconditions},  and that $\omega = 0$ and $\phi = \dd x^2$, together with $\rho_4$ and $\theta=(x^1x^2+x^4)\dd x^2 - \dd x^3$, then indeed solve the generalised invariance conditions \eqref{condition1gH} and \eqref{condition2gH}. Interestingly, this solution works for Euclidean signature of the worldsheet only.

\paragraph{Acknowledgements.} A.Ch. would like to thank the European Institute for Sciences and their Applications (EISA), 
and in particular 
Ifigenia Moraiti and George Zoupanos. 
The work of L.J. was supported by the Croatian Science Foundation under the project IP-2014-09-3258, as well as by the H2020 Twinning project No. 692194,  "RBI-T-WINNING". 
T.S. is grateful to the CNRS for according a delegation and an attachment to beautiful IMPA for one semester and 
equally to IMPA and its staff for their hospitality during this stay.
We all acknowledge support by COST (European Cooperation in Science
and Technology) in the framework of the action MP1405 QSPACE.

\end{document}